\documentclass[twocolumn, a4paper, prl, amsmath, amssymb, amsfonts,superscriptaddress]{revtex4}
\usepackage{graphicx}
\usepackage{dcolumn}
\usepackage{bm}
\usepackage{subfigure}
\usepackage{color}

\begin{document}


\title{Multi-ultraflatbands tunability and effect of spin-orbit coupling in twisted bilayer transition metal dichalcogenides}

\author{Zhen Zhan}
 \affiliation{Key Laboratory of Artificial Micro- and Nano-structures of Ministry of Education and School of Physics and Technology, Wuhan University, Wuhan 430072, China}
\author{Yipei Zhang}%
\affiliation{Key Laboratory of Artificial Micro- and Nano-structures of Ministry of Education and School of Physics and Technology, Wuhan University, Wuhan 430072, China}%
\author{Pengfei Lv}%
\affiliation{Key Laboratory of Artificial Micro- and Nano-structures of Ministry of Education and School of Physics and Technology, Wuhan University, Wuhan 430072, China}%
\author{Hongxia Zhong}%
\affiliation{Key Laboratory of Artificial Micro- and Nano-structures of Ministry of Education and School of Physics and Technology, Wuhan University, Wuhan 430072, China}%
\author{Guodong Yu}
\affiliation{Key Laboratory of Artificial Micro- and Nano-structures of Ministry of Education and School of Physics and Technology, Wuhan University, Wuhan 430072, China}
\affiliation{Institute for Molecules and Materials, Radboud University, Heijendaalseweg 135, NL-6525 AJ Nijmegen, The Netherlands}
\author{Francisco Guinea}
\affiliation{Fundaci\'on IMDEA Nanociencia, C/Faraday 9, Campus Cantoblanco, 28049 Madrid, Spain}
\author{Jos\'{e} \'{A}ngel Silva-Guill\'{e}n}
\email{josilgui@gmail.com}
\affiliation{Key Laboratory of Artificial Micro- and Nano-structures of Ministry of Education and School of Physics and Technology, Wuhan University, Wuhan 430072, China}
\author{Shengjun Yuan}
\email{s.yuan@whu.edu.cn}
\affiliation{Key Laboratory of Artificial Micro- and Nano-structures of Ministry of Education and School of Physics and Technology, Wuhan University, Wuhan 430072, China}

\begin{abstract}
 Ultraflatbands that have been theoretically and experimentally detected in a bunch of van der Waals stacked materials showing some peculiar properties, for instance, highly localized electronic states and enhanced electron-electron interactions. 
 In this Letter, using an accurate tight-binding model, we study the formation and evolution of ultraflatbands in transition metal dichalcogenides (TMDCs) under low rotation angles. 
 We find that, unlike in twisted bilayer graphene, ultraflatbands exist in TMDCs for almost any small twist angles and their wave function becomes more localized when the rotation angle decreases. 
Lattice relaxation, pressure and local deformation can tune the width of the flatbands, as well as their localization. 
 Furthermore, we investigate the effect of spin-orbit coupling on the flatbands and discover spin/orbital/valley locking at the minimum of the conduction band at the K point of the Brillouin zone. 
 The ultraflatbands found in TMDCs with a range of rotation angle below $7^\circ$, may provide an ideal platform to study strongly correlated states.  
\end{abstract}

\maketitle

\textit{Introduction.}--Stacked van der Waals layered systems provide an ideal platform to modulate the electronic properties of their parent materials via different degrees of freedom, for example, the rotation angle \cite{Bistritzer2011,Brihuega2012}. 
One of the most interesting phenomenon in these twisted two-dimensional (2D) materials is the formation of flatbands. 
Recently, it has been discovered that, in the so-called magic-angle twisted bilayer graphene, a flatband forms near the Fermi level and strongly correlated states, for instance, a Mott insulating behavior and unconventional superconductivity, arise from such flatband \cite{Cao_in,Cao_sup}. 
This generated an intensive investigation on this matter in order to identify bilayer systems that present this kind of electronic properties and that could be used as an ideal platform to study many-body interaction physics.\cite{Kerelsky2019,Xie2019,Wolf2019,Wu2018}. 

The engineering of the quantum states of matter is an active area of the experimental and theoretical research on modern condensed matter physics. 
In two-dimensional crystals and van der Waals materials, such controllable engineering can be realized by means of the rotation angle, pressure, strain or local deformation \cite{roldan2017theory,amorim2016novel}. 
For instance, the modification of the magic angle value of twisted bilayer graphene has been realized by application of a uniaxial strain \cite{Carr2018}. 
In fact, atomically thin 2D materials are particularly suited for strain engineering.
For example, single-particle bound states can be created and confined by strain at the center of bubbles in monolayers of TMDCs \cite{Chirolli2019}. 
A strain superlattice can lead to bands which describe a topological insulator \cite{Cazalilla2014}.
As a designing parameter, the interlayer coupling can be tuned by variable local stackings to tailor the electronic properties of van der Waals materials \cite{Zhang2017,Liu2014,Puretzky2016}. Recently, flatbands have been both theoretically predicted and experimentally observed in  transition metal dichalcogenides (TMDCs) \cite{Naik2018,Naik2020,Fleischmann2019,venkateswarlu2020electronic,Zhang2020,Wu2018,Wang2019,Wetal19,Jin2019,Regan2020,Sung2020,Tang2020}. 
The question arises how other tuning parameters (for instance, moir\'{e} effect, the pressure, local deformation and the spin-orbit coupling) engineer the ultraflatbands and their novel properties of twisted bilayer TMDCs. 

In this Letter, we use an accurate \textit{ab initio} tight-binding Hamiltonian \cite{Fang2015,Zhen2020} to investigate the engineering of the ultraflatbands in twisted bilayer TMDCs with low rotation angles. The tight-binding propagation method is adopted to calculated the electronic properties of the moir\'e supercell, especially for samples with tiny rotation angles \cite{yuan2010modeling,Zhen2020NC}. 
We construct the twisted bilayer TMDCs by starting from a 2H stacking ($\theta=0^\circ$) and rotating the top layer with an angle $\theta$ with respect to the bottom layer around an atom site \cite{SI,Zhen2020}. 
The moir\'e pattern has $C_3$ symmetry with the threefold rotation axis perpendicular to the TMDCs plane. As illustrated in Ref.~\onlinecite{SI}, in the supercell we can distinguish three different high-symmetry stackings (AB, $\rm B^{S/S}$ and $\rm B^{Mo/Mo}$).   
The band structure of monolayer transition metal dichalcogenides can be described by a tight-binding Hamiltonian consisting of eleven orbitals, the $d$ orbitals from Mo and the $p$ orbitals from the S \cite{cappelluti2013,Roldan2014,roldan2014electronic}.
The generalization to the bilayer case (including the twisted bilayer system) is done by adding an interlayer hopping term of the $p$ orbitals of the chalcogen between adjacent layers to a two single-layer Hamiltonian \cite{cappelluti2013,Fang2015,Zhen2020}.
Tight-binding models are quite useful for the investigation of large-scale complicated systems and for the systematic study of local strain effect on the electronic properties of 2D materials.     

\begin{figure}[t!]
\centering
\includegraphics[width = 0.48\textwidth]{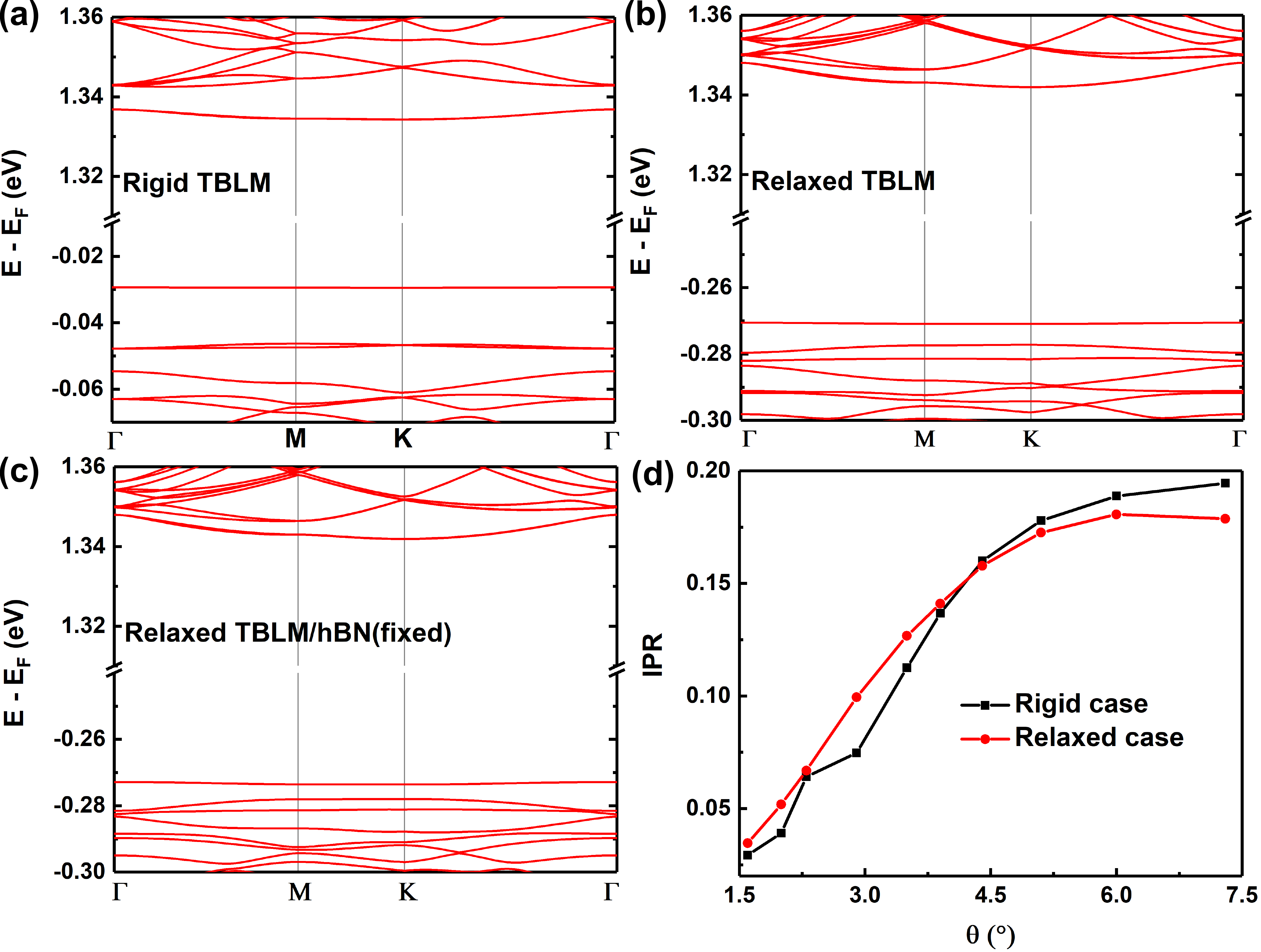}
\caption{The band structure of (a) rigidly twisted bilayer $MoS_2$, (b) relaxed TBLM and (c) relaxed TBLM/hBN with $\theta=2.0^\circ$. The hBN layer is fixed in a flat configuration. (d) Variation of the inverse participation ratio (IPR) with the rotation angle. Spin-orbit coupling is not included in the calculation.}
\label{relax}
\end{figure}

\textit{Rotation angle.}--Via exact diagonalization of the \textit{ab initio} tight-binding Hamiltonian of the twisted bilayer $\rm MoS_2$ (TBLM), we carefully study the evolution of band structures under different rotation angles from $\theta = 7.3^\circ$ to $\theta = 1.6^\circ$. When the rotation angle approaches $0^\circ$, for instance, for rigidly twisted bilayer $\rm MoS_2$ with $\theta=2.0^\circ$ (see Fig.~\ref{relax}(a)), multiple energy-separated ultraflatbands emerge at the valence band (VB) edge, which are localized at the $\rm B^{S/S}$ region \cite{Naik2020,Fleischmann2019}.
Moreover, an ultraflatband forms at the conduction band (CB) edge and its wave function is localized at the $\rm B^{Mo/Mo}$ region \cite{SI}. 
These ultraflatbands resemble the quantized energy levels of bound states of a particle in a potential well, indicating that the electron suffers a strong and deep effective moir\'{e} potential \cite{Naik2020,Fleischmann2019}. The bandwidth (energy difference between the $\Gamma$ and K points in the valence band edge) undergoes a drastic change from 30 meV to nearly zero as the rotation angle decreases from $7.3^\circ$ to $1.6^\circ$ \cite{SI}.
The ultraflatbands also occur at the valence band edge of twisted bilayer $\rm MoSe_2$, $\rm WS_2$ and $\rm WSe_2$ as well \cite{SI}, which is consistent with a recent experimental result \cite{Zhang2020}. There are two reasons that can explain the flattening of bands. 
i) The first is trivial, the growing size of the moir\'{e} unit cell shrinks the moir\'e Brillouin zone (MBZ). 
Since the MBZ is smaller than the original Brillouin zone (BZ), the energy bands simply fold into the MBZ and could cause the band flattening.
ii) The evolution of the interlayer interaction due to the formation of a moir\'{e} structure can also cause band flattening while resulting in a non-trivial modulation of the electronic properties. 
It was discovered both theoretically and experimentally that in twisted bilayer systems, the electronic states of the flatbands are highly localized at the $\rm B^{S/S}$ or $\rm B^{Mo/Mo}$ high-symmetry stacking regions in the real space, trapped by the effective periodic moir\'{e} potential and forming networks analogous to arrays of quantum dots \cite{Naik2018,Naik2020,Fleischmann2019}.

\textit{Relaxation.}--In a real twisted bilayer system which is composed of various high-symmetry stackings, since the stackings have different binding energies, the lattice structure spontaneously relaxes to achieve an energetically favorable structure \cite{Naik2018}. In the reconstructed moir\'e supercell, different high-symmetry stacking patterns have different interlayer distances. Such variation is expected to affect the electronic properties of the supercell. 
In order to study this effect we perform relaxations in two different realistic systems, a free-standing sample and another which is deposited on a hexagonal boron nitride (hBN) substrate. We perform the relaxations using the LAMMPS \cite{plimpton1995fast,lammps} package with the intralayer Stilliner-Weber (SW) potential \cite{jiang2015parametrization} and the interlayer Lennard-Jones (LJ) potential \cite{rappe1992uff}. 
Then the interlayer hoppings are reconstructed according to the positions of the atoms. 
Our calculations show that the calculated TB band structure based on the relaxed structure agrees well with the DFT result obtained from \textsc{Siesta} \cite{SI,soler2002siesta,artacho2008siesta,perdew1996generalized,artacho1999linear,grimme2006semiempirical,cuadrado2012fully}. 
As shown in Fig. \ref{relax}(b) and (c), the first band in the valence band is still quite flat in the TBLM with $\theta=2.0^\circ$, whereas with a reduced energy separation from other valence flatbands. Importantly, the spatial localization of the ultraflatbands changes from the $\rm B^{S/S}$ site to the AB site \cite{SI}. The states are still localized at the AB site even if the atoms are kept fixed in the in-plane \cite{SI}. Therefore, the underlying physical mechanism responsible for the formation of the ultraflatband in the valence band is the combination of the interlayer van der Waals interaction and the in-plane strain \cite{Naik2020}. Moreover, a flatband with a larger bandwidth forms at the conduction band, which is still localized in the $\rm B^{Mo/Mo}$ site.



To compare the localization of wave functions in the TBLM, we calculate the inverse participation ratio (IPR), which in a tight-binding model with $N$ orbitals is defined as \cite{odagaki1986electronic,Huang2019}:
\begin{equation}
IPR = \displaystyle\sum_{i=1}^{N}|a_i^{\alpha}|^2/ (N \displaystyle\sum_{i=1}^{N}|a_i^{\alpha}|^4),
\end{equation}
where $a_i^\alpha$ is the amplitude for the eigenstate $\alpha$ at the site $i$. 
The evolution of the IPR of the states at K in the VB edge is plotted in Fig. \ref{relax}(d). 
As the rotation angle becomes smaller, the IPR decreases monotonically, showing an evidence of electron localization. 
The lattice relaxation has a different effect on the localization in the TBLM supercell depending on the rotation angle. That is, it makes the states more localized in large angle samples, whereas weaken the localization for systems with small angles.

\begin{figure}[t]
\centering
\includegraphics[width = 0.48\textwidth]{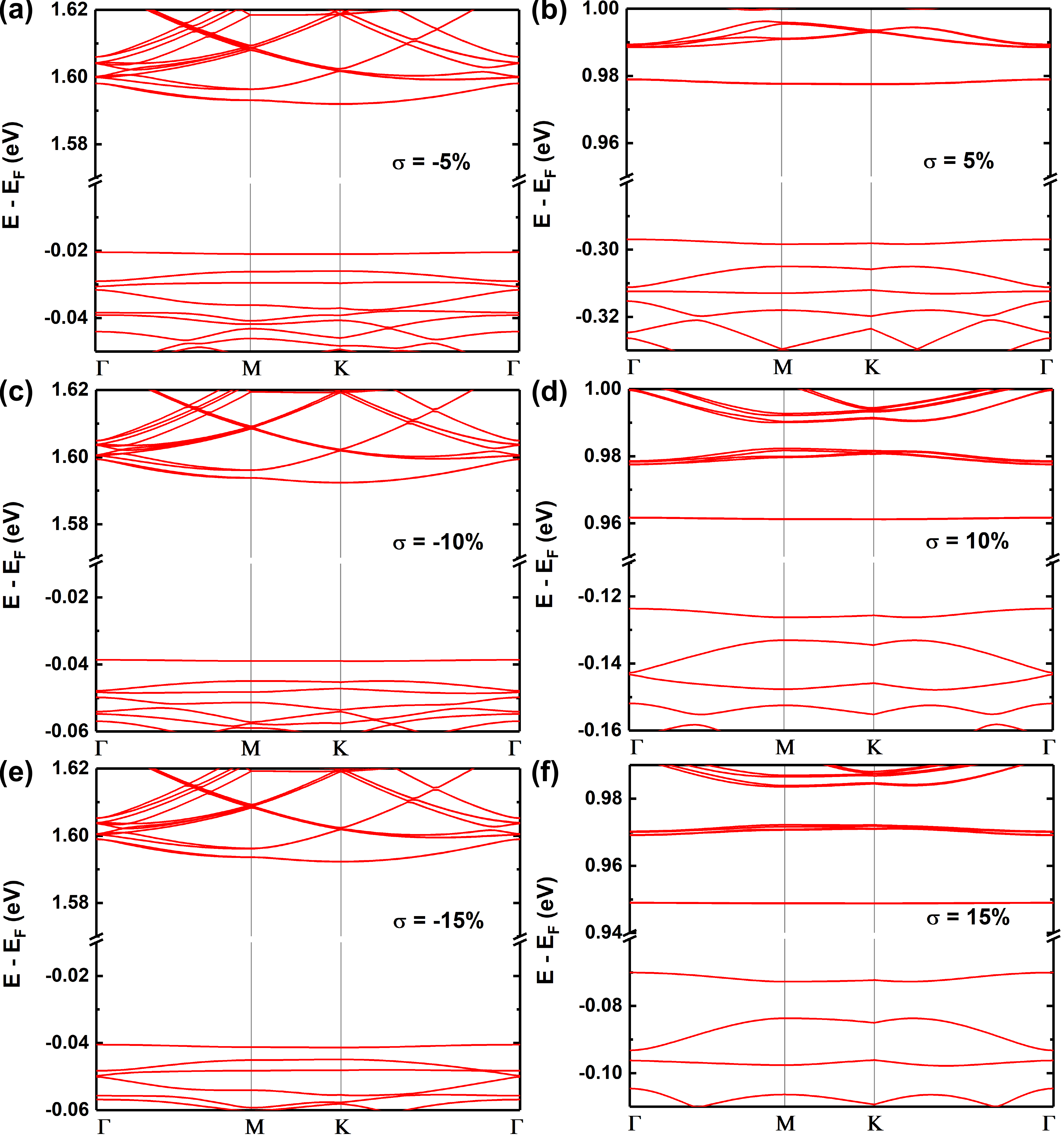}
\caption{The band structure (without SOC) of relaxed TBLM with $\theta = 2.0^\circ$ under different vertical compression.}
\label{compression}
\end{figure}

\textit{Compression.}--The possibility of modifying the ultraflatbands and their associated correlated physics in twisted bilayer TMDCs by the application of uniaxial compression is investigated.
In Fig. \ref{relax} we can see that, for TBLM with $\theta=2.0^\circ$, the ultraflatbands emerge in both CB and VB edges. 
Compression in the direction perpendicular to the bilayers is implemented in terms of $\sigma=1-(d'_{2H}/d_{2H})$ where $d'_{2H}$ is the distance between repeating two monolayer units and $d_{2H}=12.29$ \AA \; is the distance at zero compression. 

If we pull apart the layers (negative compression), isolated ultraflatbands near the CB and VB edges go deeper into the CB and VB, respectively and disappear, whereas the top VB ultraflatband is robust. 
More interestingly, as the positive compression increases, the layers come closer and the effective interlayer coupling strength increases, which creates more energy-separated ultraflatbands in both CB and VB edges. The top flatband in the VB has a larger energy separation from other bands in the systems suffering a higher positive compression.  
Such multiple energy-separated ultraflatbands is similar to the DFT result in Ref. \onlinecite{Naik2020}. 
Compression does not change the localization of the ultraflatbands \cite{SI}. 
Moreover, a progressive closure of the band gap is obtained as the compression increases, which indicates that a metallization of twisted TMDCs may occur at compression larger than 15\%. 
Such metallization could be investigated in an experimental setup with the application of pressure on TBLM higher than 56 GPa \cite{SI,bandaru2014effect,zhao2015pressure}.

\begin{figure}[t]
\centering
\includegraphics[width = 0.48\textwidth]{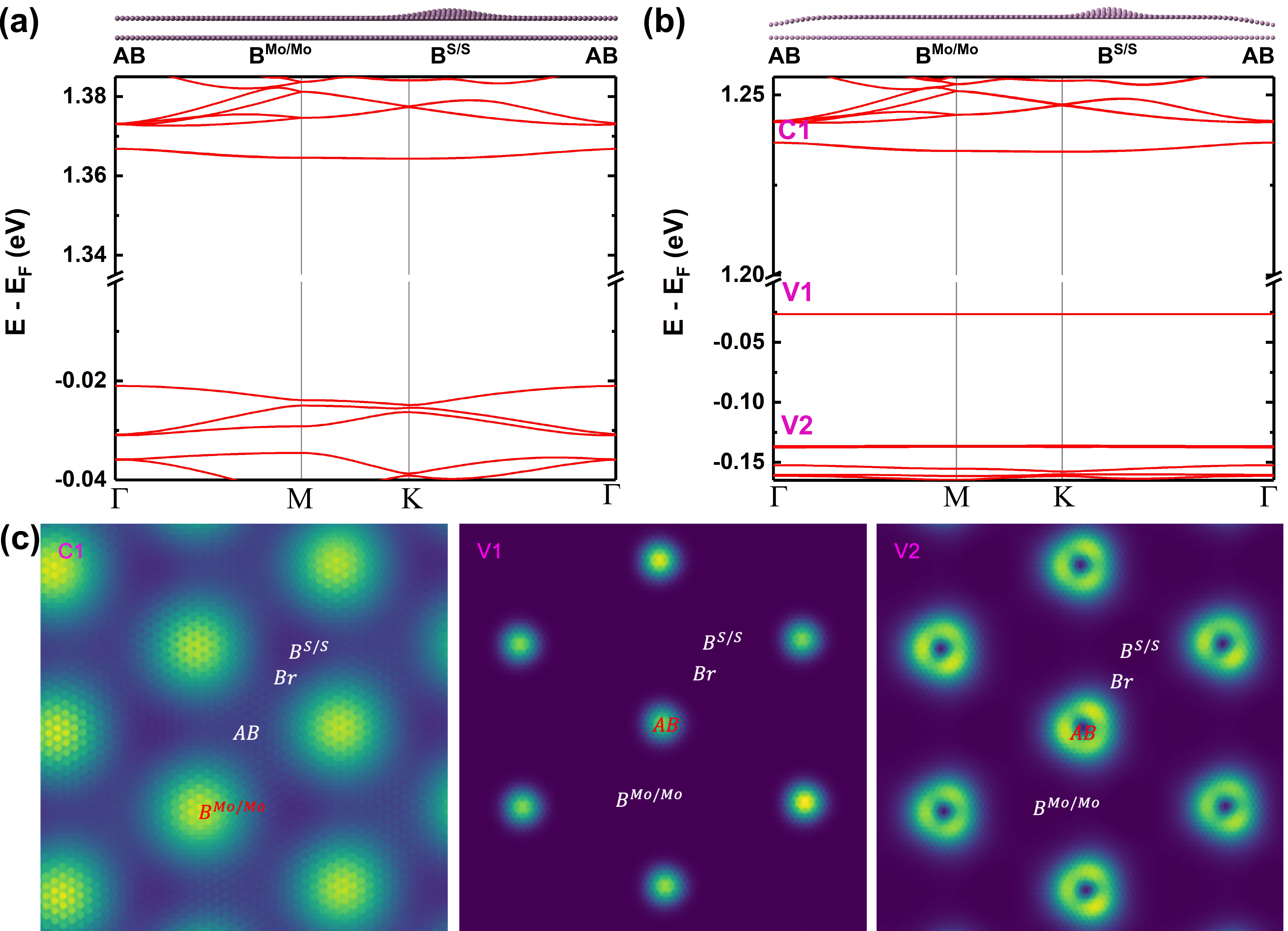}
\caption{The band structure (without SOC) of $2^\circ$ rigidly twisted bilayer $\rm MoS_2$ with local deformation at (a) the $\rm B^{S/S}$ region and (b) both the AB and $\rm B^{S/S}$ regions. The insets are the side view of the atomic model along the direction of the three high-symmetry stacking (AB, $\rm B^{Mo/Mo}$ and $\rm B^{S/S}$). The local deformation is realized by implementing a Gaussian-type bubble at the $\rm B^{S/S}$ or AB regions and with the center located at the $\rm B^{S/S}$ or AB site, respectively. (c) The calculated local density of states mapping with energies of the CB and VB edges labeled in (b).}
\label{z_direction}
\end{figure}

\textit{Local deformation.}--Specific strain textures can be produced in a system by applying an external strain, indenting with nanopillars patterned in a substrate or stacking one layer on another lattice-mismatched layer \cite{Guinea2010,Branny2017,Plantey2014,Li2015,Liu2018}. 
Such deformation has a significant influence on the electronic properties of the system. 
For instance, spatially tailored pseudo-magnetic fields were detected in graphene-based devices \cite{Liu2018}. 
How will the local deformation affect the ultraflatbands in twisted bilayer TMDCs is still unclear. 
In this part, a Gaussian-type bubble with a radius $R=26$ \AA\; is created at a high-symmetry stacking region (AB or $\rm B^{S/S}$) of the TBLM with $\theta=2^\circ$. 
The in-plane separation between the high-symmetry stacking sites is 90 \AA. 
The center of the bubble is located at either the AB or $\rm B^{S/S}$ site (see Fig. \ref{z_direction}) and the maximum out-of-plane displacement, $h_{max}$, at both AB and $\rm B^{S/S}$ are $0.05d_{2H}$. 
The bubble has a height-over-radius ratio $h_{max}/R=0.04$, in which the in-plane lattice deformation can be neglected safely. 
We only concentrate on the effect of interlayer couplings and, for simplicity, we will keep the same in-plane hopping value that independently of the local deformation and also do not relax the system.   

In Fig. \ref{z_direction}(a), the multiple ultraflatbands that localize at the $\rm B^{S/S}$ region are destroyed upon the presence of the bubble. 
When the interlayer distance increases inside the bubble, the interlayer interaction decreases, which kills the ultraflatband in the VB edge. 
After, we can generate a concave bubble with the same shape at the AB region (see the top of Fig. \ref{z_direction}(b)). 
Interestingly, as seen in Fig. \ref{z_direction}(b), multiple energy-separated ultraflatbands form again in the VB edges. 
When the height of the bubble increases, more ultraflatbands appear in the VB \cite{SI}. 
Different from the moir\'e pattern without local deformation, as shown in Fig. \ref{z_direction}(c), the new ultraflatband states are localized at the AB region. 
Differently, the local deformation in both AB and $\mathrm{B^{S/S}}$ regions have minor changes to the ultraflatband in the CB edge and its localization. 
All in all, the local deformation is remarkably efficient to tune the ultraflatband as well as its localization.

\begin{figure}[t]
\centering
\includegraphics[width = 0.48\textwidth]{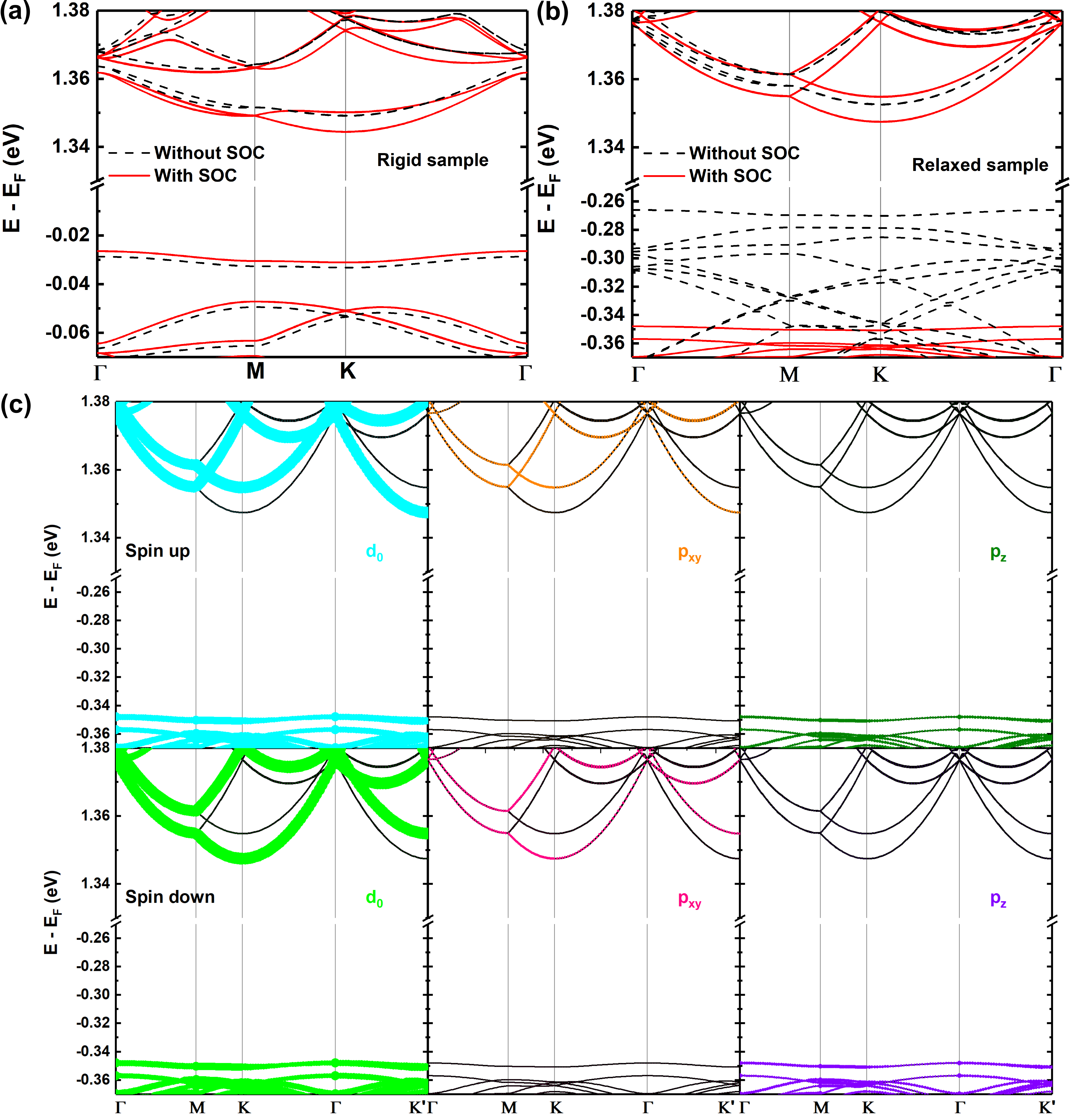}
\caption{The band structures with and without SOC of twisted bilayer $\rm MoS_2$ with rotation angles $\theta = 3.5^\circ$ in the (a) rigid case and (b) relaxed case. 
(c) The band structure (with SOC) and orbital weight of the relaxed TBLM with $3.5^\circ$. The thickness of the bands represents the orbital weight with the $d$ character ($d_2 = d_{x^2 - y^2},d_{xy}$, $d_1 = d_{xz},d_{yz}$, $d_0 = d_{3z^2-r^2}$) refers to the Mo atom $4d$ orbitals and $p$ character ($p_{xy} = p_x,p_y$) refers to S atom $2p$ orbitals. The orbital weight of $d_2$ and $d_1$ are not shown here. The sum orbital weight of $d_0$, $p_{xy}$ and $p_z$ is up to 97\%.}
\label{soc}
\end{figure}

\textit{Spin-orbit coupling.}--Transition metal dichalcogenides, in particular, single-layer TMDCs, have strong spin-orbit coupling (SOC) and broken inversion symmetry which lead to opposite spin polarization on different valleys. The locked spin and valley pseudospin gives rise to rich valley physics and makes TMDCs promising materials for next generation optoelectronic applications \cite{suzuki2014valley,scuri2020electrically}. For non-twisted bilayer TMDCs, no valley-dependent spin polarization is detected due to the presence of the inversion symmetry. Such symmetry can be broken by applying an external electric field perpendicular to the bilayer \cite{wu2013electrical}. In this section, we investigate the effect of SOC on the band structure of twisted bilayer $\rm MoS_2$, where the mirror symmetry is broken by the rotation angle. 
In the tight-binding model, the effect of  SOC is well captured by doubling the orbitals and including an on site term $\displaystyle\sum_{\alpha} \lambda_{\alpha} \mathbf{L}\cdot \mathbf{S}$ in the Hamiltonian, where subscript $\alpha$ stands for the type of atom\cite{Roldan2014,Fang2015,silva2016electronic}.  
The band structure with (red solid line) and without spin-orbit coupling (dashed black line) are shown in Fig \ref{soc}(a) and (b) for unrelaxed and relaxed samples with $\theta=3.5^\circ$, respectively. 

From these figures we observe that, upon rotation, as a consequence of breaking the mirror symmetry, spin degeneracies are lifted along the $\Gamma$--K--M path, but the time reversal invariant points $\Gamma$ and M remain spin degeneracy.
Furthermore, opposite to the single-layer case, the effect of SOC is more significant in the conduction band for the twisted bilayer system.
In the monolayer, the conduction band edge, at the $K$ point, has a $d_0$ character \cite{cappelluti2013}. For these states, $\langle \mathbf{L}\cdot \mathbf{S} \rangle = 0$, and the effect of the spin-orbit requires processes of second order in perturbation theory \cite{ochoa2013}. 
The states at the $K$ point of MoS$_2$ have a contribution to the flatbands studied here \cite{SI}. These states have a significant S $p_x , p_y$ character, where the splitting due to the spin-orbit coupling is a first order process. 
Meanwhile, at the conduction band edge, the K valley has the largest spin splitting and shrinks when the rotation angle decreases. 
Such shrink in the relaxed samples is smoother than the rigid case \cite{SI}. Focusing on the valence band edges, we find that the ultraflatband at the valence band edge is doubly spin-degenerate in the whole BZ. As shown in Fig. \ref{soc}, there is no spin splitting at the VB edge. That is highly consistent with the DFT result \cite{kumari2020engineering}.

We further calculate the orbital weight for the band structure of TBLM with $\theta=3.5^\circ$. States at the bottom of the conduction band mainly consist of $p_{xy}$ and $d_0$ orbitals, and the two in-equivalent valleys are spin-locked. That is, the spin up states at valley K are degenerate with spin down states at valley K', and vice-versa. 
That is different from the monolayer case where the Q point has the largest band splitting with a spin-polarization.
The ultraflatband at the valence band edge is mainly composed of $d_0$ and $p_z$ orbitals and is indeed spin-degenerate. These bands are located away from the K point of the $\rm MoS_2$ non-twisted bilayer (consists of $d_2$ and $p_{xy}$ orbitals), and thus show a reduced SOC splitting \cite{SI}.
In the CB an ultraflatband is formed and its localization is at the $\rm B^{Mo/Mo}$  region. 
Such band is almost composed solely by $d_0$ orbital while it has a finite contribution from $p_{xy}$ orbitals. 
This indicates a different origin of the ultraflatbands near the conduction band. 
The electronic states at the conduction band edge are nearly decoupled from each other for two different layers, as there is little interlayer interaction for the $\rm B^{Mo/Mo}$ stacking. That explains why the relaxation and local deformation do not change the localization of the VB edge state. 
This property might allow independent manipulation of electronic states in conduction band minimum for the two different layers.

\textit{Conclusions.}--We have studied the evolution of the band structure of twisted bilayer MoS$_2$. 
We found that as the rotation angle decreases, the band width decreases monotonically and the flatband wave functions become more localized as well.
When the rotation angle is below a certain value, flatbands start to emerge at the valence band and multiple energy-separated flatbands form at the conduction band. 
Furthermore, compression and mechanical strain are effective methods to tune the flatbands and their localization in real space. 
Finally, we analyzed the orbital composition near the band edge in the presence of SOC. 
We show that the states in the conduction band edge are spin-polarized, and the polarization is opposite in K and K'. Therefore, the effect of spin-orbit coupling should be taking into account when developing simpler models for this kind of systems. The physical mechanism responsible for the formation of flatbands in the CB and VB are different. 
We also found that the electronic states at conduction band edge are independent for the two layers according to their orbital contribution. 
Note that if one wishes to tune the flatband significantly using the rotation angle, it is better to suppress the structural relaxation in the system. Ultraflatbands are also detected in other twisted bilayer TMDCs, for instance, $\rm MoSe_2$, $\rm WS_2$ and $\rm WSe_2$ \cite{SI}, which indicates that twisted TMDCs could be used as an ideal platform for the understanding of correlated behaviors.\\

\textit{Acknowledgments}--This work was supported by the National Science Foundation of China under Grant No. 11774269. G Yu acknowledges a support from the China Postdoctoral Science Foundation (Grant No. 2018M632902). F. G. acknowledges support by funding from the European Commision, under the Graphene Flagship, Core 3, grant number 881603, and by the grants NMAT2D (Comunidad de Madrid, Spain),  SprQuMat and SEV-2016-0686, (Ministerio de Ciencia e Innovación, Spain). Numerical calculations presented in this paper have been performed on the supercomputing system in the Supercomputing Center of Wuhan University. J.A.S.G. acknowledges the computer resources at Marigold and technical support provided by SCAYLE (RES-FI-2020-2-0040).


\bibliography{flatband2}

\end{document}